\documentclass[prb,aps,twocolumn,amsmath,amssymb,superscriptaddress,affilletter]{revtex4-1}

\usepackage{graphicx}
\usepackage{tabularx}
\usepackage[T1]{fontenc}
\usepackage[cp1250]{inputenc}
\usepackage[rightcaption]{sidecap}
\newcommand{\beq}{\begin{equation}}
\newcommand{\eeq}{\end{equation}}

\begin{document}

\title{Revealing the nature of excitons in liquid exfoliated monolayer tungsten disulphide}

\author{Ł.~Kłopotowski}
\email[Corresponding author: ]{lukasz.klopotowski@ifpan.edu.pl}
\affiliation{Institute of Physics, Polish Academy of Sciences, Al. Lotników 32/46, 02-668 Warsaw, Poland}

\author{C.~Backes}          \affiliation{Centre for Research on Adaptive Nanostructures and Nanodevices (CRANN), Trinity College Dublin, Dublin 2, Ireland}
\affiliation{School of Physics, Trinity College
	Dublin, Dublin 2, Ireland}
\affiliation{Applied Physical Chemistry, University of Heidelberg, Im Neuenheimer Feld 253, 69120 Heidelberg, Germany}

\author{A.~A.~Mitioglu}          \affiliation{LNCMI, CNRS-UJF-UPS-INSA, Grenoble and Toulouse, France}               \affiliation{Institute of Applied Physics,
	Academiei Str. 5, Chisinau, MD-2028, Republic of Moldova}

\author{V.~Vega-Mayoral} \affiliation{Department for Complex Matter, Jozef Stefan Institute, Jamova 39, 1000 Ljubljana, Slovenia}
\affiliation{ Jozef Stefan International Postgraduate School, Jamova 39 , 1000 Ljubljana, Slovenia}

\author{D.~Hanlon}          \affiliation{Centre for Research on Adaptive Nanostructures and Nanodevices (CRANN), Trinity College Dublin, Dublin 2, Ireland}
\affiliation{School of Physics, Trinity College
	Dublin, Dublin 2, Ireland}

\author{J.~N.~Coleman}          \affiliation{Centre for Research on Adaptive Nanostructures and Nanodevices (CRANN), Trinity College Dublin, Dublin 2, Ireland}
\affiliation{School of Physics, Trinity College
	Dublin, Dublin 2, Ireland}

\author{V.~Y.~Ivanov}
\affiliation{Institute of Physics, Polish Academy of Sciences, Al. Lotników 32/46, 02-668 Warsaw, Poland}

\author{D.~K.~Maude}       \affiliation{LNCMI, CNRS-UJF-UPS-INSA, Grenoble and Toulouse, France}

\author{P.~Plochocka}      \affiliation{LNCMI, CNRS-UJF-UPS-INSA, Grenoble and Toulouse, France}

\begin{abstract}
Transition metal dichalcogenides hold promise for applications in novel optoelectronic devices. There is therefore a need for materials that can be obtained in large quantities and with well understood optical properties. In this report, we present a thorough photoluminescence (PL) investigations of monolayer tungsten disulphide obtained via liquid phase exfoliation. As shown by microscopy studies, the exfoliated nanosheets have dimensions of tens of nanometers and thickness of 2.5 monolayers on average. The monolayer content is about 20\%. Our studies show that at low temperature the photoluminescence is dominated by excitons localized on nanosheet edges. As a consequence, the PL is strongly sensitive to environment and exhibits an enhanced splitting in magnetic field. As the temperature is increased, the excitons are thermally excited out of the defect states and the dominant transition is that of the negatively charged exciton. Furthermore, upon excitation with a circularly polarized light, the PL retains a degree of polarization reaching 50\% and inherited from the valley polarized photoexcited excitons. The studies of PL dynamics reveal that the PL lifetime is on the order of 10 ps, probably limited by non-radiative processes. Our results underline the potential of liquid exfoliated TMD monolayers in large scale optoelectronic devices.
\end{abstract}

\maketitle

\section{Introduction}

Semiconducting transition metal dichalcogenides (TMDs) form a new class of materials with potential applications ranging from gas sensors and energy storage devices to field effect transistors and novel light sources \cite{wan12,but13}. For applications in optoelectronics, monolayer TMDs are particularly well suited. Compared to bulk, their direct bandgap \cite{mak10,spl10} increases the photon emission efficiency by several orders of magnitude \cite{mak10,zha13a,zen13}, while sub-nanometer layer thickness decreases dielectric screening and results in large exciton binding energies \cite{ram12} making excitonic effects present at room temperature \cite{han15}. Recently, these properties were exploited to demonstrate the operation of an excitonic laser \cite{ye15} and a strong coupling of an exciton and a photon in a planar microcavity \cite{liu14}. Moreover, the particular band structure of monolayer TMD exhibits two inequivalent minima (valleys) at $+K$ and $-K$ corners of a hexagonal Brillouin zone. The $\pm K$ valleys constitute a binary index (pseudospin) that can be addressed with optical transitions in $\sigma^{\pm}$ polarizations \cite{zen12,mak12,cao12}, respectively, opening a way to applications in quantum information processing \cite{xia12,xu14}.

The majority of the work reported so far was done on mechanically exfoliated (cleaved) TMD monolayers. Although perfect for fundamental investigations, mechanical exfoliation is tedious, not reproducible, and clearly not scalable and thus poorly suited for applications. In order to implement TMDs in optoelectronic devices, establishing a procedure that allows to obtain large areas of monolayer material is required. One approach is to employ a bottom-up growth technique, e.~g., chemical vapour deposition (CVD). It was recently reported that CVD growth allows to obtain MoS$_2$ and WS$_2$ monolayers with areas exceeding ~1 mm$^2$ \cite{dum15,mcc16}. An alternative solution is a top-down method of liquid phase exfoliation \cite{col11,nic13}, in which a bulk powder is sonicated or sheared in an appropriately chosen solvent, which also prevents the mono- or few-layer nanosheets from reaggregation. This versatile and inexpensive technique allows to obtain large quantities of monolayer material \cite{bac14,var15,lon15,she15,yi15,bac16}, which can be then used for ink-jet printing \cite{sec15}, spray coating, forming hybrids and composites \cite{woo14}, depositing thin films \cite{may12}, and functionalization \cite{cui11}.

Liquid phase exfoliation is a novel method for producing solutions containing TMD nanosheets with a high monolayer content.\cite{bac16} However,
different monolayer extraction techniques are known to influence the optical properties of the resulting material.\cite{pei13,con14} This can be due to varying strain or concentration of defects, adsorption of foreign atoms, or different dielectric environment resulting from the exfoliation or growth procedure. Therefore, before liquid exfoliated monolayer TMDs can be used in optoelectronic devices, a detailed investigation of their optical properties is required. For liquid exfoliated TMDs, such studies are, until now, lacking. This is mostly because resultant raw dispersions after liquid exfoliation are highly polydipserse containing nanosheets with broad length and thickness distributions and hence low monolayer contents. We have recently addressed this by establishing a new size selection technique termed liquid cascade centrifugation \cite{bac16} and studied the room temperature optical response of size-selected nanosheets. This allowed us to identify a convenient procedure to produce nanosheet dispersions with increased monolayer contents that are suitable for further fundamental studies on the optical properties. Providing the understanding of the photoluminescence (PL) from liquid exfoliated nanosheets of WS$_2$ is the objective of this work.

In this report, we present detailed PL studies on liquid exfoliated monolayer WS$_2$. The evolution of the PL spectrum with temperature shows that, at 10 K, the PL is dominated by excitons localized on defects. The studies of the temporal stability of few-nanosheet PL and the splitting of the PL transition in magnetic field both point out that the localization occurs at the edges of the nanosheets. This conclusion is consistent with small size of the nanosheets, evaluated from atomic force microscopy and a spectroscopic metric. As the temperature is increased, a negatively charged trion recombination emerges showing that the liquid exfoliated nanosheets are {\em n-}doped. Upon circularly polarized excitation, the PL signal exhibits a degree of polarization as high as 50\%, showing that the valley pseudospin is largely conserved during the energy relaxation toward the localized states. Time-resolved PL studies reveal that at 10 K the PL decays on a timescale of 10 ps, which is most probably dominated by a non-radiative processes. Our studies underline the potential of the liquid exfoliated TMD nanosheets in the device optoelectronic applications, possibly involving the valley pseudospin.

\section{Methods}

Nanosheets of WS$_2$ were produced by sonication of bulk WS$_2$ powder in aqueous sodium cholate solution. For details of the liquid exfoliation procedure see the Supplementary Information and Ref. \cite{bac16}. After subsequent size-selection by centrifugation to enrich the dispersion in monolayers, the WS$_2$ nanosheets were transferred to an aqueous poly (vinyl alcohol) (PVA) solution. The transfer to PVA was necessary to prevent reaggregation in thin films and to facilitate the further optical characterization \cite{bac14}. Thus produced nanosheets were first characterized by atomic force microscopy (AFM), transmission electron microscopy (TEM), and room temperature extinction and Raman/PL measurements.

Bright field TEM imaging was performed under 200 kV on drops of diluted nanosheet dispersions deposited on holey carbon grids. Atomic force microscopy was carried out in tapping mode after depositing a drop of the dispersion on a Si/SiO$_{2}$ wafer with an oxide layer of 300 nm. Typical image sizes were in the range of 1x1 to 2x2 $\mu$m$^2$. The apparent thickness was converted to number of layers using previously elaborated step-height analysis of liquid-exfoliated nanosheets \cite{bac14,bac16}. Statistical analysis was performed by measuring the longest dimension of the nanosheet and denoting it as "length", while the perpendicular dimension was denoted as "width". The area to determine the monolayer volume fraction was estimated as "length" $\times$ "width".

Room temperature optical extinction was measured on the stock solution in a quartz cuvette. Raman and PL spectroscopy, at both low and room temperature,  was performed on a drop-casted film on a Si/SiO$_{2}$ wafer. The excitation source for steady state studies was a diode laser emitting at 532 nm. For investigations of PL dynamics, a frequency doubled output of a an optical parametric oscillator (OPO), pumped with a Ti:Sapphire laser was used, providing pulses of $\sim 300$ fs duration. As discussed in detail below, the PL measurements were carried out in two modes. In a {\em macro}-PL mode, the excitation beam was focused with a lens to a spot with a diameter of about 200 $\mu$m. In a {\em micro}-PL mode, the excitation beam was focused to a diffraction-limited spot of $\sim$ 2 $\mu$m in diameter. The PL/Raman signal was collected with the same optics as used for focusing the excitation and dispersed and detected with a CCD coupled with a monochromator. Time-resolved PL was detected with a streak camera synchronized with the Ti:sapphire laser. The overall temporal resolution was about 5 ps. Measurements at low temperatures were performed with the sample placed on a cold-finger cryostat. PL studies in magnetic field were performed with the sample placed in a helium cryostat equipped with a split-coil superconducting magnet. During the measurements, the sample remained in gas helium at 7 K and the magnetic field was applied perpendicular to the Si/SiO$_{2}$ wafer substrate.

\section{Results and Discussion}

\subsection{Room temperature characterization}
\label{sec:char}

\begin{figure*}
  \includegraphics[angle=0,width=.95\textwidth]{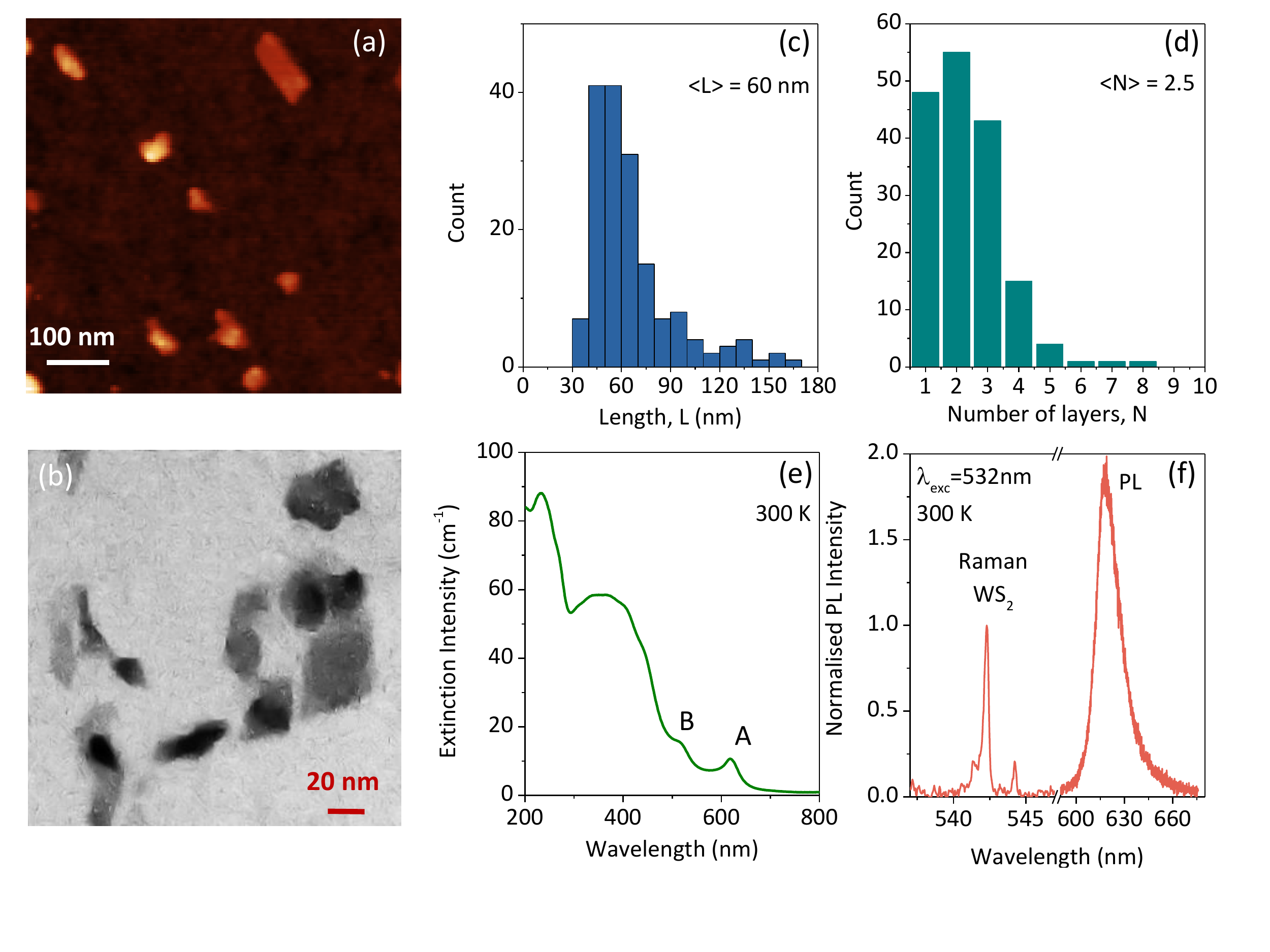}
  \caption{(a) AFM image of WS$_2$ nanosheets. (b) Close-up of a TEM image showing single WS$_2$ nanosheets. (c) Distribution of nanosheet lengths $L$ and (d) distribution of thicknesses expressed in number of monolayers $N$ obtained from the AFM analysis. (e) Room temperature extinction spectrum of of a solution of WS$_2$ nanosheets. Peaks labeled A and B correspond to the spin-orbit split excitonic transitions. (f) Room temperature PL and Raman spectra of a dried droplet containing WS$_2$ nanosheets.}
  \label{char}
\end{figure*}

We start with the discussion of the morphology of the liquid exfoliated WS$_2$ nanosheets. The AFM and TEM images shown in Figs. \ref{char}(a) and \ref{char}(b), respectively, show well-exfoliated individual nanosheets with lateral dimensions of 20-150 nm in length. AFM statistical analysis on $\sim 150$ individual nanosheets was carried out to evaluate nanosheet length and thickness distributions as well as the monolayer content. Length and thickness histograms are shown in Figs. \ref{char}(c) and \ref{char}(d), respectively. The length histogram shows that nanosheets are predominantly in the range of 40-70 nm with a mean length $<L>$ of 60 nm. We note that length from AFM is typically  overestimated by 15--20 nm due to tip broadening and pixilation effects. The apparent nanosheet thickness was converted to the number of layers using previously established step height analysis to account for residual water and surfactant/polymer which overestimates the apparent measured thickness \cite{bac14,bac16}. For this sample, the average thickness is about 2.5 monolayers (see Fig. \ref{char}(d)), while the monolayer volume fraction is about 18\%.

The extinction spectrum measured at room temperature on the obtained solution is presented in Fig. 1(e). Peaks related to the absorption by the A-- and B--excitons are clearly seen at 620 nm (2000 meV) and 515 nm (2410 meV), respectively. The measured splitting between A-- and B--excitons, resulting from the spin-orbit splitting in the conduction and valence bands, is equal to 410 meV, in very good agreement with measured \cite{gut13,pei13,zen13,con14,che14,han15} and calculated values \cite{xia12,liu13}. The extinction spectrum can be used to reliably assess mean nanosheet length and thickness due to edge and confinement effects resulting in changes to the spectral profile. Such an analysis (see the Supplementary Information and Refs. \cite{bac14,bac16} for details) gives a mean nanosheet length of 45 nm and mean thickness of 2.8 monolayers in good agreement with AFM analysis. Note that the 45 nm in mean length are more realistic than the 60 nm from the AFM (see above).

\begin{figure*}
	\includegraphics[angle=0,width=.95\textwidth]{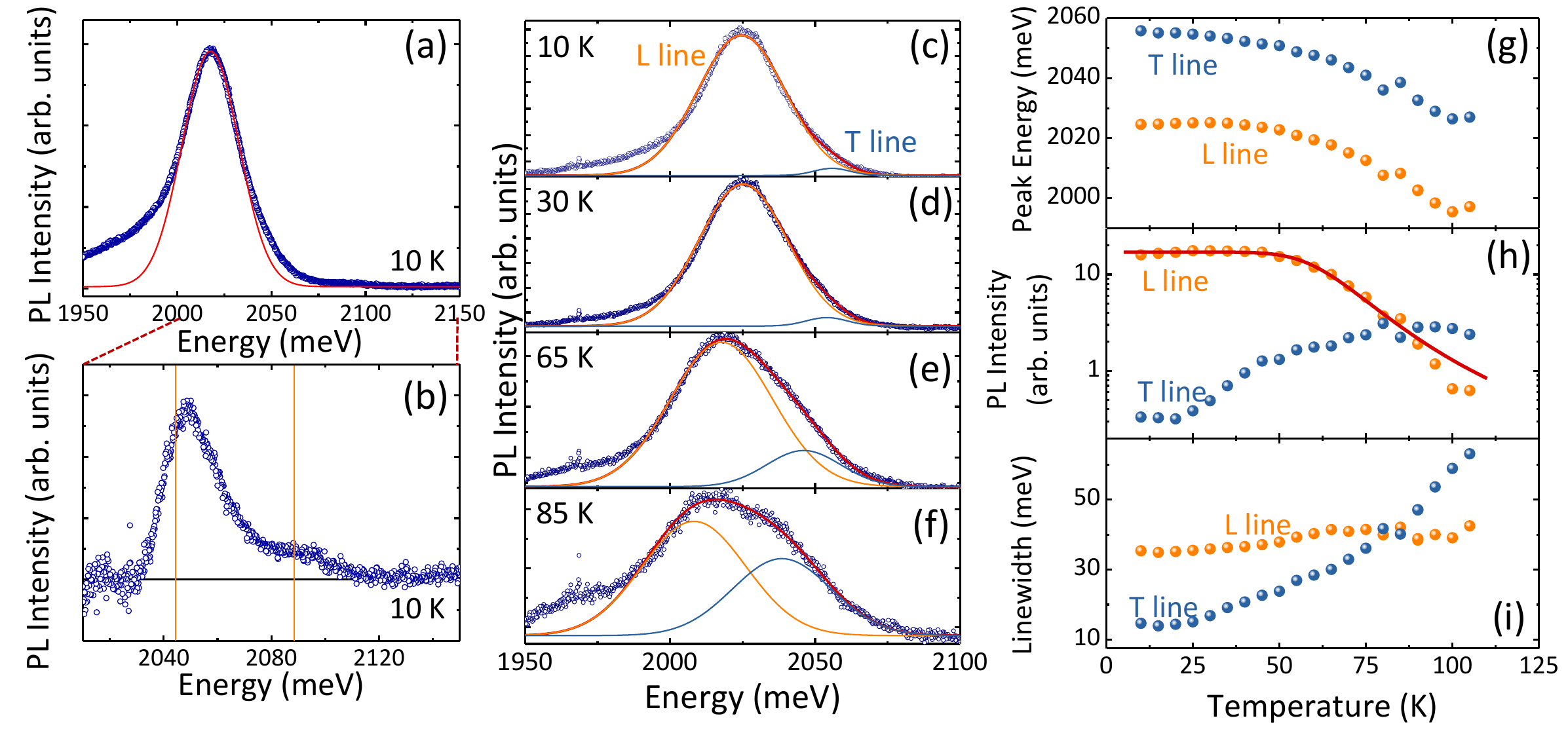}
	\caption{(a) Macro--PL spectrum of an ensemble of WS$_2$ nanosheets measured at 10 K. Red line represents a Gaussian fit to the main part of the spectrum. (b) High energy part of the PL spectrum after subtraction of the Gaussian main part. Orange vertical lines mark the PL energies of the free exciton and trion reported in Ref. \cite{ple15}. (c)-(f) PL spectra measured at various temperatures. Red lines denote fitted PL spectra comprising of two Gaussian lineshapes marked in blue (the trion PL, T line) and orange (localized exciton PL, L line). (g)-(i) Temperature dependence of the transition energies, total PL intensities, and PL linewidths for the T and L peaks extracted from fitting of the two Gaussians. The red line in (h) is a fit of the Arrhenius formula (see text).}
	\label{makroPL}
\end{figure*}

In Figure \ref{char}(f), we show the room temperature PL and Raman spectrum of a dried drop of a nanosheet solution. The PL consists of a single peak centered at about 620 nm (2000 meV) with a small shoulder extended toward longer wavelengths. As shown by tracking the PL signal during drying of the liquid drop, the PL peak corresponds to a recombination of the negatively charged trion, developing as a result of removal of electron scavenging water molecules \cite{ton13,veg15}. We assume that the PL signal originates solely from the monolayer nanosheets, as the PL intensity for a bilayer WS$_2$ is at least an order of magnitude weaker and further decreases with the number of layers \cite{zha13a,zen13,he15b}. Furthermore, the PL signal from multilayers would exhibit contributions from the transitions via the indirect bandgap. These additional peaks are expected to be redshifted by several hundred meV depending on the layer thickness with respect to the transition via the direct gap \cite{mak10,zha13a,zen13}. We do not observe any additional transitions, which confirms our assumption that the PL is due to monolayers only. At short wavelengths in Figure \ref{char}(f), the Raman peaks related to the WS$_2$ are seen. The dominant peak at $\sim 542$ nm (357 cm$^{-1}$) results from a superposition of the second order longitudinal acoustic mode 2LA(M) and the in-plane E$^1_{2g}$ mode, while the peak at $\sim 544$ nm (422 cm$^{-1}$) originates from the out-of-plane A$_{1g}$ mode \cite{ber13,zha13,zen13,mit14}. The PL/Raman intensity ratio of 3.6 measured in the liquid can also be used as quantitative measure for the monolayer content (see the Supplementary Information and Ref. \cite{bac16} for details) yielding a monolayer volume fraction of 0.21 in excellent agreement with AFM.

\subsection{Low temperature steady state photoluminescence}

We now turn to low temperature properties of the monolayer WS$_2$ photoluminescence. In Fig. \ref{makroPL}(a), we show a PL spectrum measured at 10 K. In this case, the excitation beam is focused with a lens to a spot of $\sim$ 200 $\mu$m in diameter. As noted in the Method section, we will refer to such a measurement as {\em macro}--PL. Analogously to the room temperature spectrum shown in Fig. \ref{char}(f), a single asymmetric peak is observed. The main part of the spectrum can be well fitted with a Gaussian, centered at about 2020 meV, i.e., roughly 20 meV higher in energy than at room temperature. The blueshift of the PL with decreasing temperature is expected due to the increase of the bandgap. However, if the PL peak at 10 K originated from the same transition as in room temperature, the blueshift magnitude observed here should be much larger as the bandgap is expected to open up by about 70 meV in WS$_2$ and other TMDs, when cooled in this temperature range \cite{kio12,ple15,aro15}. In order to determine the origin of the low temperature PL in the nanosheets, we subtract the fitted Gaussian from the spectrum displayed in Fig. \ref{makroPL}(a). Figure \ref{makroPL}(b) shows the high energy part of the resulting spectrum. Clearly two peaks are observed. The energies of these peaks ($\sim 2050$ meV and $\sim 2095$ meV) agree very well with the energies reported for the exciton and trion recombination at 10 K in monolayer WS$_2$ \cite{che14,han15,ple15}. We therefore conclude that the main PL peak observed at the low temperature originates from excitons localized on defect states. This conclusion is further confirmed by studies of the temperature dependence of the PL -- see Figs. \ref{makroPL}(c-f). As the temperature is increased, the intensity of the main peak (labeled L) decreases, while simultaneously another peak (labeled T and identified as the negative trion recombination) emerges at the high energy side. Over the whole investigated temperature range, the intensity of the exciton PL is negligible compared to the trion or the defect PL. We thus conclude that the nanosheets are strongly {\em n-}doped confirming the results of the room temperature experiment, which revealed the emergence of the trion upon evaporation of water \cite{veg15}. In order to gain quantitative information about the L and T transitions, we fit the temperature dependent PL spectra with two Gaussians. The temperature dependence of the peak energy, integrated intensity (peak area), and full width at half maximum (FWHM) are presented in Figs. \ref{makroPL}(g--i). As seen in Fig. \ref{makroPL}(g), both peaks redshift at the same rate with increasing temperature as a result of the shrinkage of the bandgap, confirming that both originate from the WS$_2$ nanosheets. At 10 K, the L peak is almost two orders of magnitude more intense than the T peak, but the intensities become roughly equal at about 80 K (Fig. \ref{makroPL}(g)). The temperature activated decrease of the total intensity $I(T)$ of the L peak can be fitted with the Arrhenius formula $I(T)/I(0)=1/(1+C \exp(E_A/(k_B T)))$, where $C$ is a constant and $E_A$ is the activation energy. The fitting yields $E_A = 44.3$ meV, which corresponds to 357 cm$^{-1}$ and, as seen in the Raman spectrum in Fig. \ref{char}(f), is twice the energy of the longitudinal acoustic phonon \cite{mol11,ber13}. It is therefore possible that the thermal quenching of the L peak is related to a phonon activated process. Conversely, $E_A$ can be interpreted as the average binding energy of the excitons on the defects associated with the emission in the L peak. Furthermore, while the FWHM of the T peak increases strongly with temperature, most probably as a result of the phonon induced broadening \cite{aro15}, the FWHM of the L peak remains roughly constant (see Fig. \ref{makroPL}(i)). This insensitivity to temperature shows that the FWHM in this case reflects the inhomogenous distribution of binding energies of excitons on defects. Further evidence in support of this conclusion is presented below.

\begin{figure}
  \includegraphics[angle=0,width=0.5\textwidth]{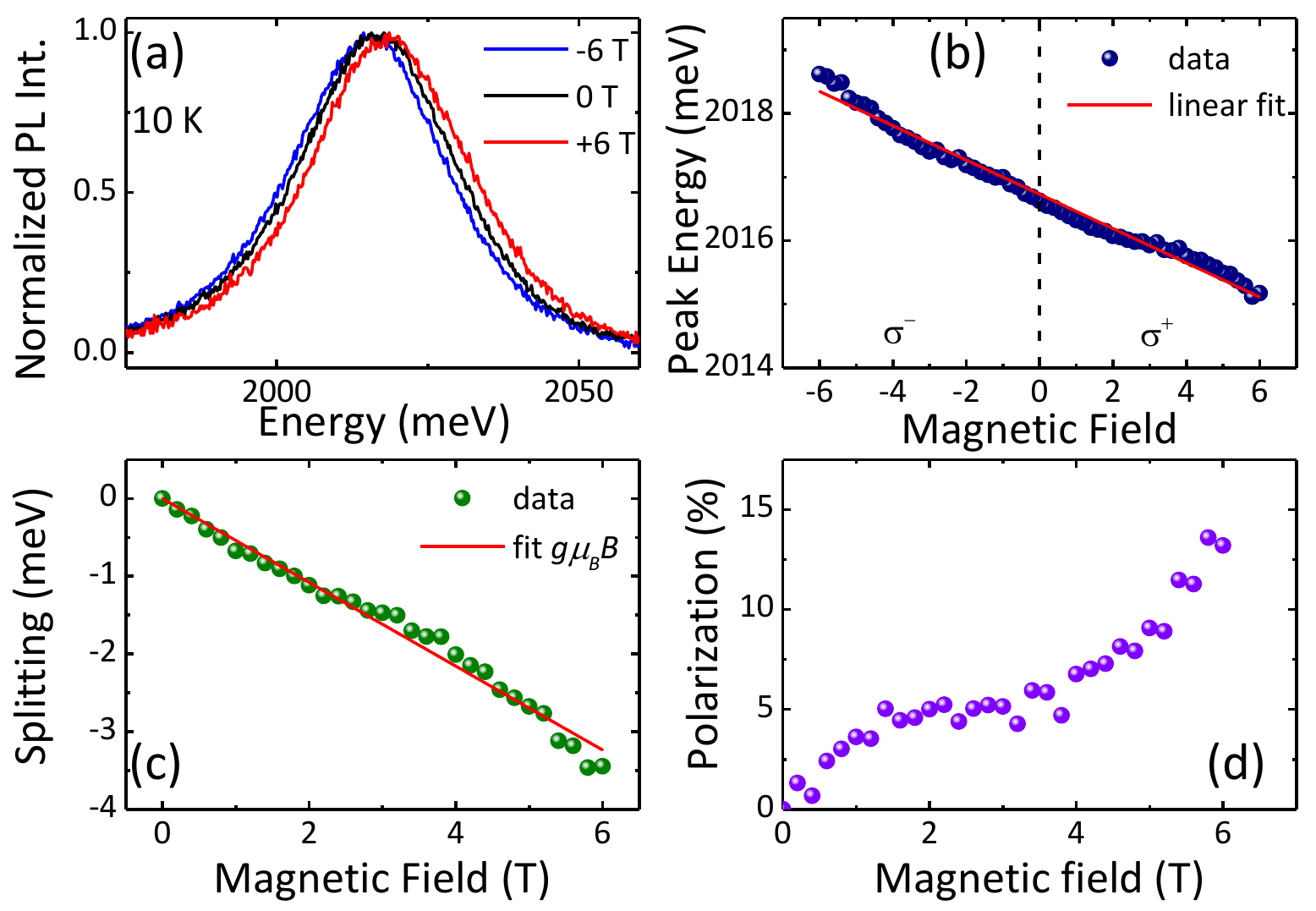}
  \caption{(a) Normalized PL spectra collected at 0 T and $\pm 6$ T with $\sigma^{+}$-polarized detection. A splitting of the PL peak is clearly resolved. (b) PL peak energies as a function of the magnetic field. The dependence of the peak energy on the magnetic field is approximately linear as demonstrated by a linear fit. The fields in  $\pm$ directions correspond to detection in $\sigma^{\pm}$ polarization, respectively. (c) The splitting $\Delta E$ between the $\sigma^{+}$- and $\sigma^{-}$-polarized PL as a function of the magnetic field $B$. The fitted linear dependence yields a $g$-factor of $-9.3$. (d) Magnetic field dependence of the circular polarization degree of the PL excited with a linearly polarized light.}
  \label{mag}
\end{figure}

The identification of the low temperature PL peak as originating from localized excitons is also supported by measurements of the macro--PL at 7 K in magnetic field. Figure \ref{mag}(a) shows normalized PL spectra measured at 0 T and at $\pm6$ T, corresponding to detection in $\sigma^{\pm}$ polarizations. A splitting of the PL peaks is clearly resolved. For quantitative analysis, the spectra were fitted with single Gaussians. Thus obtained peak energies are plotted as a function of the magnetic field in Fig. \ref{mag}(b) and showing a roughly linear dependence. The splitting of the peaks calculated as $\Delta E = E_+ - E_-$, where $E_{\pm}$ are the PL peak energies in $\sigma^{\pm}$ polarizations is plotted in Fig. \ref{mag}(c). We fit the magnetic field dependence of  the splitting as $\Delta E(B) = g \mu_{B} B$, where $\mu_{B}$ is the Bohr magneton. The fitted $g$-factor is $-9.3\pm 0.1$. This is an important result, since the $g$-factors for exciton and trion peaks evaluated for both exfoliated and CVD grown MoS$_2$, MoSe$_2$, WSe$_2$, and WS$_2$ are universally equal to about $-4$ \cite{li14,mac15,sri15z,aiv15,mit15,sti15,mit16}. The value of $g=-4$ results from the orbital magnetic moments of the $d$-orbitals of the transition metal ions, which construct the electronic states in the vicinity of the band extrema at the $\pm K$ points \cite{li14,mac15,sri15z,aiv15,mit15,sti15,mit16}. However, the universality concerns only the interband transition that conserve the carrier quasi--momentum. For strongly localized excitons, the quasi--momentum is not a good quantum number and the rule is relaxed. The origin of the increased splitting in magnetic field for localized excitons is unknown and requires studies, which are beyond the scope of the present work. We demonstrate this effect here with the purpose of supporting the identification of the PL from liquid exfoliated monolayers as originating from localized excitons.

The macro--PL spectra shown in Figs. \ref{makroPL} and \ref{mag} are collected without the use of microscopic techniques. It demonstrates that the optical properties of TMDs can be easily investigated on large areas containing liquid exfoliated nanosheets as long as the monolayer content is high enough to provide a strong PL signal. A natural trade--off is that the signal originates from an ensemble and thus suffers from inhomogenous broadening due to different size, strain, and dielectric environments of the nanosheets. In fact, the different influences of these environments can occur on length scales larger than a typical spot in a macro--PL measurement, and, as a result, spectra collected at different location of the sample can exhibit slightly different intensities, energy positions, and linewidths.

\begin{figure}[b]
  \includegraphics[angle=0,width=0.5\textwidth]{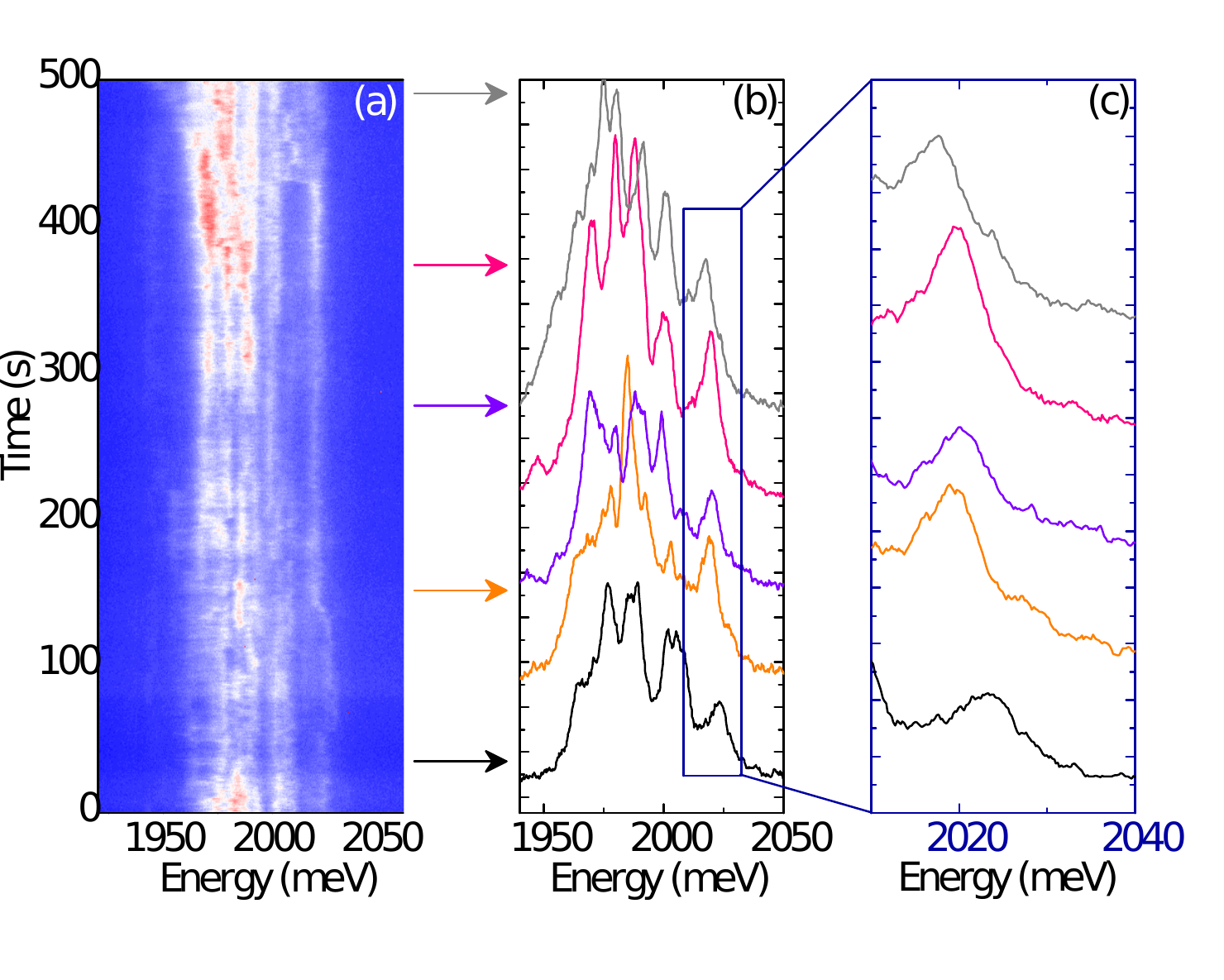}
  \caption{(a) Photoluminescence intensity plot as a function of time. Spectral wandering and blinking of the PL signal is visible. PL spectra taken at the moments marked by arrows are shown in (b). The close-up of a PL line presumably corresponding to a single nanosheet is shown in (c). }
  \label{tele}
\end{figure}

We now turn to the results of {\em micro}--PL experiments, where the excitation beam is focused with a microscope objective to a spot with a diameter of about 2 $\mu$m. These studies allow to access the PL of individual nanosheets. In Fig. \ref{tele}(a), we show a PL map, collected at 5 K, that demonstrates the temporal changes of the PL signal on a timescale of seconds. In this case, the PL signal is collected from an area with a relatively small number of monolayer nanosheets. Cross sections of the map taken at different times marked by arrows are presented in Fig. \ref{tele}(b). It is seen that the PL spectrum consists in fact of several peaks. The PL signal is unstable, showing blinking and spectral wandering characteristic for states highly sensitive to the environment. We attribute the time fluctuations of the PL positions and intensities to randomly fluctuating electrostatic environment of the nanosheets. These fluctuations result in changes in the PL spectrum due to a Stark effect analogously to PL spectra of single self-assembled quantum dots \cite{bes02}. A close-up of a single PL line is shown in Fig. \ref{tele}(c). The FWHM of this line is only about 5 meV, almost an order of magnitude smaller than evaluated for the ensemble -- see Fig. \ref{makroPL}(i), and a factor of $\sim$ 6 smaller than for a trion PL from a mechanically exfoliated WS$_2$ flake \cite{ple15}. These results confirm that the linewidth of the macro--PL Gaussian L peak (see Fig. \ref{makroPL}) is indeed due to a superposition of individual peaks at different energies. Thus, the FWHM of the L peak reflects the distribution of exciton binding energies on defects.

The results presented above allow us to discuss the nature of the defects on which the excitons are localized. As shown by a number of authors, the edges of WS$_2$ flakes exhibit different optical properties than the centers: at the edges, the PL peak is several times stronger and redshifted by tens of meV \cite{gut13,con14,kob15}. High resolution PL mapping revealed that the PL emission from the edges is not due to the recombination of neutral or charged excitons, but originate from another, deeply bound complex giving rise to narrow and temporally unstable PL peaks \cite{kat14}, similar to the ones shown in Fig. \ref{tele} This allows us to conclude that the edge states of the nanosheets are responsible for the exciton localization and for the low temperature PL signal. This conclusion is consistent with the relatively small size of the liquid exfoliated WS$_2$ nanosheets. Indeed, assuming an aspect ratio of 2 \cite{bac14}, mean length of the long edge of 45 nm (see Section \ref{sec:char}), and the in-plane lattice constant of 3.2 \AA \cite{zhu11}, we find that $\sim 4$\% of the tungsten atoms are located at the nanosheet edge. In fact, the absorbance studies of liquid exfoliated nanosheets of MoS$_2$ \cite{bac14} and WS$_2$ (not shown) revealed that the edge regions exhibiting different properties from the nanosheet centers are about 10 nm wide, i.~e.~, much wider that one lattice constant.

Distinct photoluminescence from edge states was recently discovered for monolayer WSe$_2$. Specifically, the edges were shown to contain quasi-zero dimensional centers emitting single photons \cite{cha15,he15,kop15,sri15,ton15}. This observation proved full quantization of motion of the excitons bound on these states. The zero dimensional character make these excitons extremely sensitive to the fluctuations of local electrostatic environment, which results in spectral wandering of the PL transitions due to a Stark effect \cite{cha15,he15,kop15,sri15,ton15} -- an effect analogous to the one presented in Fig. \ref{tele}. Moreover, the splittings of those quasi-zero-dimensional excitons localized on edge defects exhibit a much larger $g$-factor compared to the neutral exciton or trion \cite{cha15,he15,kop15,sri15}. In fact, a reported survey of several of such emitters reveal an average value of $g = -8.7$ \cite{he15}, very close to the $g = -9.3$ that we obtain from an analysis of the ensemble PL. Taking all the above arguments into account, it seems reasonable to conclude that the edge states responsible for the L peak in the low temperature PL spectra are indeed associated with quasi-zero dimensional centers. However, while there is a strong circumstantial evidence for this identification, an unequivocal confirmation would require photon correlation measurements.

\subsection{Valley polarization}

As mentioned in the Introduction, one of the most exciting properties of monolayer TMDs is the possibility to create carriers in specific valleys using circularly polarized excitation. The valley polarization can be monitored via the circular polarization of the photoluminescence \cite{zen12,mak12,cao12}. This property is inherent to free excitons and trions, whose momentum is a good quantum number. As discussed above, this is not the case for localized excitons and, consequently, the polarization of defect PL is usually much weaker \cite{mak12,cao12,kio12,jon13,wan14}.

\begin{figure}
  \includegraphics[angle=0,width=0.5\textwidth]{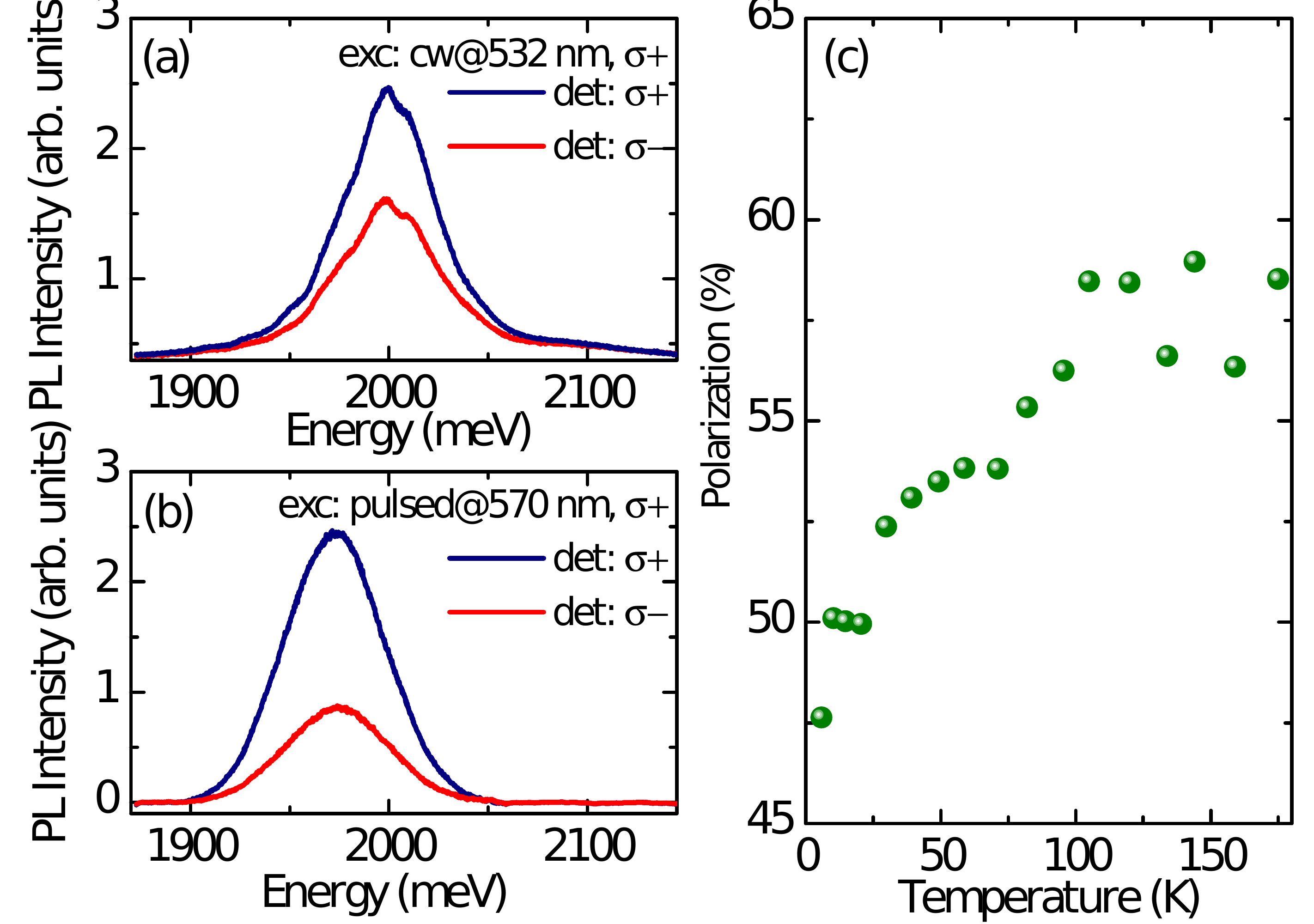}
  \caption{PL spectra  measured at 5 K on a large ensemble of WS$_2$ nanosheets excited with a $\sigma^+$-polarized laser beam continuously at 532 nm (a) and with OPO pulses at 570 nm (b). Blue (red) curves denote PL spectra co- (counter-)polarized with the excitation. (c) Temperature dependence of the PL polarization.}
  \label{pol}
\end{figure}

In Fig. \ref{pol}(a), we show micro--PL spectra measured at 5 K on an area containing a large number of monolayer nanosheets and, therefore, the narrow peaks due to single emitters are mostly smeared out. The excitation in this case is $\sigma^+$--polarized and we detect the PL in two circular polarizations. As seen in Fig. \ref{pol}(a), the co-polarized PL signal is significantly stronger than the counter-polarized one. We define the degree of circular polarization as $P = (I^+-I^-)/(I^++I^-)$, where $I^{\pm}$ are the total PL intensities in $\sigma^{\pm}$ polarizations. For the spectra in Fig. \ref{pol}(a), we obtain $ P \approx 25$\%. Figure \ref{pol}(b) shows the micro--PL spectra excited with a $\sigma^+$--polarized OPO pulse at 570 nm. In this case, the PL has approximately a Gaussian lineshape and is slightly redshifted with respect to the continuous excitation case. The origin of this redshift is not known and we speculate that it can be due to changes in the PVA conformation under the femtosecond pulse, which in turn affects the nanosheets via strain and modifications of the dielectric environment. The polarization in this case exceeds 50\%. The different values of $P$ hint at a specific excitation wavelength dependence, which is a subject of a separate study. Here, we point out that approaching the A--exciton resonance, $P$ increases as for free exciton PL \cite{mak12,kio12}, showing that the circular polarization of the localized exciton PL in Figs. \ref{pol}(a) and \ref{pol}(b) is inherited from the photoexcited, valley polarized free excitons. More specifically, spin-valley coupling resulting from the lack of inversion symmetry in monolayer TMDs\cite{xia12} dictates that $\sigma^{\pm}$--polarized excitation creates only spin-up excitons in the $\pm K$ valley. Although the valley index is not defined for localized carriers as demonstrated by time-resolved Kerr rotation studies \cite{yan15,yan15nl}, nonzero $P$ shows that the spin is preserved from the valley polarized free excitons during the energy relaxation to the localized states.

The temperature dependence of $P$ is shown in Fig. \ref{pol}(c). The excitation wavelength is in this case 550 nm. As the temperature is increased from 5 to 180 K, the polarization increases slightly from about 50\% to about 60\%. As pointed out by the Kerr rotation studies, spin lifetime of localized electrons is expected to quickly decrease with temperature. Such an effect is not observed for the localized excitons studied here and we interpret the increase of $P$ with temperature as resulting from an increased contribution of the trion PL in the total PL signal -- see discussion of Fig. \ref{makroPL} above. Finally, we note that the origin of the circular polarization analyzed in Fig. \ref{pol} is different from the magnetic field induced polarization shown in Fig. \ref{mag}(d). In the latter case, since the excitation is linearly polarized, the circular polarization results from preferential population of the lower energy state, optically active in $\sigma^+$ polarization \cite{mac15}.

\subsection{Photoluminescence dynamics}

\begin{figure}[!h]
\includegraphics[angle=0,width=0.5\textwidth]{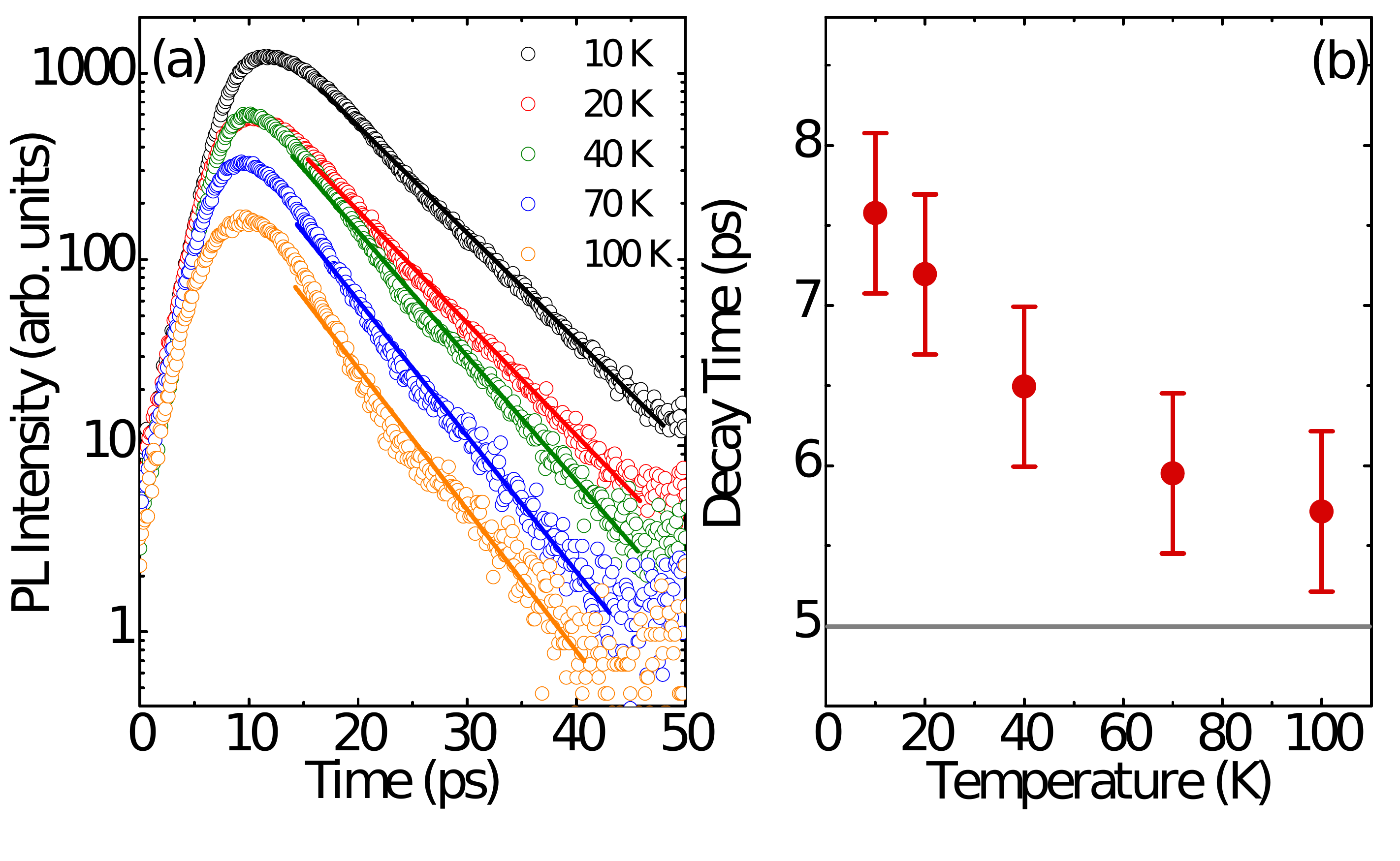}
\caption{(a) Points: PL time decays measured for various temperatures. Lines: fitted monoexponential decays. (b) PL decay times as a function of temperature. Grey line denotes the resolution limit.}
\label{decays}
\end{figure}

Before the implementation in optoelectronic devices is achieved, a detailed knowledge of light emission dynamics is necessary. In Fig. \ref{decays}(a), we show the decays of the energy-integrated PL, measured in the temperature range between 10 and 100 K. At 10 K, the PL decays is less than 10 ps and slightly decreases with increasing the temperature. In Fig. \ref{decays}(b), we plot the PL lifetimes extracted by fitting a single exponential decay to the measured temporal dependence. The lifetime shortens from almost 8 ps at 10 K to less than 6 ps, close to our temporal resolution, at 100 K. As demonstrated for quantum wells \cite{fel87} and predicted for monolayer TMDs \cite{pal15}, for free excitons whose recombination is governed by a radiative process, a strong increase of the lifetime with temperature is expected. The relatively weak temperature dependence shown in Fig.~\ref{decays} support the conclusion that the recombining excitons are indeed localized. Shortening of the PL lifetime with temperature could be due to increased influence of the trion emission as discussed in the context of Fig. \ref{makroPL}. Notably, the lifetimes of defect- and edge-related PL in TMDs are usually substantially longer.\cite{wan14,he15,sri15} We therefore conclude that the PL decays observed in Fig. \ref{decays} are controlled by non-radiative processes. 

\section{Conclusions}

In conclusion, we have presented a thorough investigation of the origin of the PL from liquid exfoliated monolayers of WS$_2$. Analysis of the temperature dependence of the PL lineshape allows us to conclude that at low temperature the PL is dominated by
the recombination of localized excitons. Two additional results allow to specify the nature of the localized states: (i) the PL splitting in magnetic field is more than twice larger than for free excitons in monolayer TMDs and (ii) the analysis of the temporal instability of the few-nanosheet PL shows that the PL spectrum consists of narrow peaks related to individual centers strongly sensitive to the environment. These two results together with small nanosheet size (of tens of nm as measured by AFM) and in analogy to the data published by other groups\cite{gut13,con14,kat14,kob15} allow us to conclude that the exciton localization occurs at the nanosheet edges. Furthermore, the PL exhibits circular polarization reaching 50\%, induced by the circularly polarized excitation and inherited from the valley-polarized free excitons via spin-valley coupling. Time-resolved studies reveal PL lifetimes on the order of 10 ps, probably controlled by non-radiative processes. Our studies show that liquid exfoliated monolayers of WS$_2$ exhibit excellent optical properties that allow to study subtle effects related to the nanosheet band structure. Crucially, we show that these effects can be accessed without microscopic techniques and that large scale devices can be easily fabricated. We believe that this work will pave the way to design and study of optoelectronic devices based on liquid exfoliated TMDs.

\section*{Acknowledgements} This work was partially supported by Investissements d'Avenir under the program ANR-11-IDEX-0002-02, IDEX program EMERGENCE, project BLAPHENE, ANR JCJC project milliPICS, the Region Midi-Pyr\'en\'ees under contract MESR 13053031 and STCU project 5809. ŁK acknowledges the support of Institut National des Sciences Appliqu\'{e}es Toulouse. JNC and CB acknowledge funding from the European Union Seventh Framework Program under grant agreement n°604391 Graphene Flagship. VVM acknowledges Marie Curie ITN network "MoWSeS" (Grant No. 317451).


\providecommand{\newblock}{}

\end{document}